\documentclass[journal]{IEEEtran}

\usepackage{cite}
\usepackage{url}
\usepackage{graphicx}
\usepackage{color}
\usepackage{bm}
\usepackage{cite}
\usepackage{amsmath,amssymb,amsfonts}
\usepackage{algorithmic}
\usepackage{graphicx}
\usepackage{textcomp}
\usepackage{algorithmic}
\usepackage{algorithm}
\usepackage{array}
\usepackage[caption=false,font=normalsize,labelfont=sf,textfont=sf]{subfig}
\usepackage{stfloats}
\usepackage{url}
\usepackage{verbatim}
\usepackage{enumitem}
\usepackage{tikz}
\usepackage{bm}
\usepackage{threeparttable,booktabs}
\usepackage{blkarray}
\usepackage{amsmath}
\usepackage{multirow}

\usepackage{amsmath}
\usepackage{amsthm,amsfonts}
\usepackage{pgfplots}

\usepackage{tikz}
\usepackage[customcolors]{hf-tikz}
\usepgfplotslibrary{colorbrewer}
\usetikzlibrary{%
  3d,
  arrows,%
  shapes.misc,
  shapes.arrows,%
  chains,%
  matrix,%
	fit,%
  positioning,
  scopes,%
  decorations.pathmorphing,
  shadows,%
  decorations.markings,
  shapes.geometric,
  arrows.meta,bending,
	patterns,
	intersections, 
	pgfplots.fillbetween
}

\usepackage{tikz-3dplot}
\usepackage[pdfusetitle]{hyperref}
\hypersetup{
    colorlinks,
    linkcolor={red!50!black}, 
    citecolor={blue!50!black},
    urlcolor={blue!50!black}
}

\usepackage{cellspace}
\DeclareMathAccent{\ring}{\mathalpha}{operators}{"17}

\renewcommand{\Re}{\operatorname{Re}}	
\renewcommand{\Im}{\operatorname{Im}}	

\newcommand{\ie}{\textit{i.e.}\/, }

\newcommand{\cf}{\textit{cf.}\/, }

\colorlet{dpurple}{blue!50!red}
\colorlet{dblue}{blue!50!black}
\colorlet{dgreen}{green!50!black}
\colorlet{dred}{red!50!black}
\colorlet{dyellow}{yellow!50!black}
\colorlet{dorange}{orange!50!black}
\definecolor{metal}{RGB}{218,165,32}
\definecolor{diel}{RGB}{1,165,32}
\definecolor{antenna}{RGB}{100,150,162}
\definecolor{breg}{rgb}{0.2,0.6,0.8}%
\definecolor{preg}{rgb}{0.8,0.2,0.2}%
\definecolor{reg}{RGB}{218,165,32}

\graphicspath{{figures/}{tikz/}}

\makeatother

\begin{document}
\title{Time-Reversal Characteristic Modes of Lossy Reciprocal Structures}

\author{Chenbo Shi, Jin Pan, Xin Gu, Shichen Liang, and Le Zuo
\thanks{Manuscript received May 16, 2026. (\textit{Corresponding author: Jin Pan.})}
\thanks{Chenbo Shi, Jin Pan, Xin Gu, and Shichen Liang are with the School of Electronic Science and Engineering, University of Electronic Science and Technology of China, Chengdu 611731, China (e-mail: chenbo\_shi@163.com; panjin@uestc.edu.cn; xin\_gu04@163.com; lscstu001@163.com).}
\thanks{Le Zuo is with the 29th Research Institute of China Electronics Technology Group Corporation (e-mail: zorro1204@163.com).}
}


\maketitle

\begin{abstract}
A time-reversal characteristic-mode decomposition is developed for reciprocal lossy electromagnetic structures. The formulation is built on a transmit--receive interpretation of reciprocity: the far-field pattern radiated by a mode determines the time-reversed incident field that is optimally matched to couple energy back into that same mode. This physical picture leads to an antilinear characteristic-mode equation whose solutions remain radiation-power orthogonal even in the presence of material loss, lossy loading, or matched absorption. As a result, the modal expansion coefficients directly represent the radiated-power contributions of the corresponding modes and avoid the singular biorthogonal normalization that may arise in nonnormal classical characteristic-mode expansions. Equivalent formulations are derived in the scattering-operator, T-matrix, and method-of-moments (MoM) frameworks, thereby connecting external wave-channel descriptions with current-space and port-excitation descriptions. The proposed modes reduce to classical characteristic modes in the lossless limit. Numerical examples involving a lossy two-sphere system and a loaded folded antenna demonstrate the radiation-power orthogonality, modal-expansion stability, and power interpretability of the proposed decomposition near exceptional points, where classical characteristic-mode expansions become singular or lose their radiated-power meaning.
\end{abstract}


\section{Introduction}

\IEEEPARstart{C}{haracteristic} mode theory (CMT) is a valuable modal-decomposition tool whose central idea is to represent the external response of an electromagnetic structure through a set of intrinsic modes determined by the structure itself \cite{ref_CM_Garbacz,ref_CM_Harrington}. Such a modal description clarifies current distributions, radiation patterns, and scattering properties, and it provides a unified language for modal selection, excitation design, mode suppression, and wideband modal tracking. As a result, the range of applications benefiting from CMT has expanded rapidly~\cite{ref_CM_APP1,ref_CM_APP2,ref_CM_APP3,ref_CM_APP4}. For lossless systems, CMT has developed into a mature theoretical framework~\cite{ref_CM_Die1,ref_CM_Die2,ref_CM_Die3,ref_CM_YlaOijala_generalized,ref_CM_Hu_cyl1,ref_CM_Hu_cyl2}. Gustafsson \emph{et al.} unified the MoM-based current definition with the scattering-based definition \cite{ref_Unified_I,ref_Unified_II}. In the scattering framework, characteristic modes can be formulated as a standard spectral problem, yielding a clear physical picture in which a mode is reproduced by the structure up to a complex proportionality factor. In the current framework, the same set of modes is expressed as a weighted spectral problem, namely a generalized eigenvalue problem. The equivalence between the two formulations further reveals an elegant algebraic connection between the scattering and current descriptions.

However, once dissipative mechanisms such as material loss, lossy loading, or matched absorption are introduced, one of the defining virtues of characteristic modes in lossless systems---radiation-power orthogonality---is lost~\cite{ref_Unified_II,ref_CM_Lossy0,ref_CM_Lossy1,ref_CM_Orth}. This loss has two immediate consequences. First, the projection formula used in lossless CMT can no longer be applied directly to obtain modal expansion coefficients. Existing lossy-CMT formulations therefore resort to biorthogonal expansions, or equivalently reaction-based orthogonality, between the left and right eigenvectors of the generally nonnormal scattering operator to recover algebraic modal-expansion coefficients~\cite{ref_Unified_II,ref_CM_Orth}. Second, even after such coefficients are obtained, they no longer admit a direct radiated-power interpretation. This treatment, while formally valid in many cases, conceals a fundamental fragility: the stability and well-posedness of a biorthogonal expansion rely on the existence of a complete eigenvector set and on a nonsingular biorthogonal normalization. These assumptions are precisely the ones that become questionable under strong nonnormality, modal coalescence, or in the vicinity of exceptional points (EPs). In such regimes, a large modal coefficient need not represent a large radiated contribution; it may instead be an artifact of an ill-conditioned or even singular expansion. Thus, although biorthogonality formally restores an expansion formula, it does not restore the radiation-power interpretation that makes characteristic modes useful for engineering design. This limitation weakens the modal-attribution capability of classical CMT for lossy structures and can lead to misleading conclusions in mode-based analysis and design.

A natural remedy is to seek a modal basis that remains radiation-power orthogonal even in the presence of loss. Several approaches have pursued this goal from different perspectives. A representative example is provided by Inagaki modes~\cite{ref_CM_svd}, which are obtained by applying singular value decomposition to the scattering operator or the T-matrix. This construction yields orthogonal input--output channels and therefore restores a well-defined power decomposition at the channel level. However, it does so at the price of splitting a characteristic mode into two generally different objects: an input singular vector that determines how the channel is excited, and an output singular vector that determines the far field radiated by that channel. As a result, a single modal object no longer simultaneously carries the three pieces of information central to characteristic-mode analysis: excitation matching, radiation pattern, and power contribution. For modal selection, excitation synthesis, and mode suppression, an additional pairing between the input and output singular vectors must therefore be imposed or interpreted externally. A further practical limitation is that such channels are naturally suited to plane-wave or spherical-wave scattering problems, whereas, for local excitations commonly encountered in antennas, such as port feeding, the channel definition does not directly prescribe how the excitation should be designed. Other studies have worked directly in the MoM framework. For example, \cite{ref_Radiation_modes1,ref_Radiation_modes2} constructs radiation modes from the MoM radiation matrix. Although this method yields an orthogonal radiation basis and supports power expansion, the resulting modes do not possess a clearly paired physical excitation picture. In addition, even in the lossless limit, such definitions do not reduce to the well-established classical characteristic modes.

In this work, we consider reciprocal lossy systems and seek a radiation-power-orthogonal modal definition that remains physically interpretable in the presence of loss. In a reciprocal system, the radiation pattern of a mode should not only describe how the mode radiates outward; it should also determine the external wave that is optimally matched to couple energy back into that mode. From this viewpoint, if the far-field pattern radiated by a structure is time-reversed, \ie played backward and sent back to illuminate the structure itself, and if the structure then regenerates only the original radiation pattern, this pattern can be defined as a characteristic mode. The modes obtained in this way are referred to in this paper as time-reversal characteristic modes. We show that, even in lossy structures, time-reversal characteristic modes preserve radiation-power orthogonality. Consequently, the corresponding modal expansion is stable and avoids the singular biorthogonal normalization associated with defective nonnormal eigenproblems. More importantly, the expansion coefficients retain a direct physical meaning: they quantify the contribution of each mode to the total radiated power.

We provide equivalent definitions of time-reversal characteristic modes in several computational frameworks, including the scattering operator, the T-matrix, and MoM matrices. This construction is inspired by the unified CMT framework \cite{ref_Unified_I}, which connects the scattering-based and MoM-based current formulations through a common algebraic structure. The scattering and T-matrix formulations most directly reveal the underlying time-reversal picture, whereas the MoM formulation transfers the modal definition to current degrees of freedom, port-excitation vectors, and local feeding mechanisms, thereby making the theory directly applicable to antenna problems. Together, these formulations provide a unified and computable description of time-reversal characteristic modes for radiation and scattering problems. Moreover, the proposed definition is constructed in a form closely resembling classical CMT: in the lossless limit, it naturally reduces to the same form as classical lossless characteristic modes, reflecting the time-reversal symmetry of lossless systems. This structural resemblance allows computational strategies developed for classical CMT to be adopted with minimal modification. For example, the scattering or T-matrix formulation can be integrated into commercial solvers that support plane-wave scattering \cite{ref_OA_code1,ref_OA_code2,ref_Iterative,ref_Sdyadic}, whereas the MoM formulation can be combined with matrix-free MoM and iterative eigensolvers for electrically large structures \cite{ref_CM_MLFMM}. 

It is worth emphasizing that the definition of time-reversal characteristic modes relies on reciprocity rather than passivity. Therefore, the proposed theory applies not only to lossy structures, but can also be extended to reciprocal active or gain-assisted structures. Passivity imposes additional constraints on the channel strength and eigenvalue distribution. As long as reciprocity holds, however, the physical picture of paired transmission and reception remains valid. In this sense, the present work provides a general radiation-power-orthogonal modal-analysis framework for lossy, loaded, and reciprocal active electromagnetic structures.

\section{Theory of Time-Reversal Characteristic Modes}

Consider a finite obstacle $\Omega$, and assume monochromatic time-harmonic fields with the time convention $\exp(\mathrm{j}\omega t)$. We begin with the external scattering-operator description, in which a general incident field is expanded over the complete continuum of incident plane-wave components. For the component propagating in direction $\hat{\boldsymbol{k}}$, let $\boldsymbol{A}(\hat{\boldsymbol{k}})$ denote the complex amplitude of the incident electric field. The scattering operator maps the incident-field amplitude distribution $\boldsymbol{A}(\hat{\boldsymbol{k}})$ to the outgoing far-field amplitude $\boldsymbol{F}(\hat{\boldsymbol{r}})$, defined through the large-$r$ asymptotic relation
\begin{equation}
\boldsymbol{F}(\hat{\boldsymbol{r}})
=
r e^{\mathrm{j}kr}\boldsymbol{E}_s(\boldsymbol{r})
=
-\frac{4\pi}{\mathrm{j}k}
\int \mathrm{d}\hat{\boldsymbol{k}}\,
\bar{\boldsymbol{S}}(\hat{\boldsymbol{r}},\hat{\boldsymbol{k}})
\cdot
\boldsymbol{A}(\hat{\boldsymbol{k}})
\label{eq:farfield_scattering}
\end{equation}
where $\mathrm{d}\hat{\boldsymbol{k}}$ denotes the measure on the unit sphere, and the observation point is written as $\boldsymbol{r}=r\hat{\boldsymbol{r}}$. $\bar{\boldsymbol{S}}(\hat{\boldsymbol{r}},\hat{\boldsymbol{k}})$ is the scattering dyadic, defined as in~\cite{ref_Sdyadic}. In this work, we consider reciprocal structures. Accordingly, the scattering dyadic satisfies the Lorentz reciprocity relation \cite{ref_Sca}
\begin{equation}
\bar{\boldsymbol{S}}(\hat{\boldsymbol{r}},\hat{\boldsymbol{k}})
=
\bar{\boldsymbol{S}}^{\mathrm{T}}(-\hat{\boldsymbol{k}},-\hat{\boldsymbol{r}}).
\label{eq:reciprocity_sdyadic}
\end{equation}

Classical scattering-based characteristic modes can be written as the standard spectral problem
\begin{equation}
\mathcal{S}\boldsymbol{F}_n
=
t_n \boldsymbol{F}_n
\label{eq:classical_cm_operator}
\end{equation}
or, equivalently, in integral form as \cite{ref_Unified_I,ref_Sdyadic}
\begin{equation}
\int \mathrm{d}\hat{\boldsymbol{k}}\,
\bar{\boldsymbol{S}}(\hat{\boldsymbol{r}},\hat{\boldsymbol{k}})
\cdot
\boldsymbol{F}_n(\hat{\boldsymbol{k}})
=
t_n \boldsymbol{F}_n(\hat{\boldsymbol{r}}).
\label{eq:classical_cm_integral}
\end{equation}
Here, $\boldsymbol{F}_n$ is a right eigenfunction of the operator $\mathcal{S}$. In lossless reciprocal structures, this eigenproblem admits a radiation-power-orthogonal choice of modes. With dissipation, $\mathcal{S}$ generally becomes nonnormal, so a projection based only on right eigenfunctions is no longer available. The classical expansion is therefore constructed through the biorthogonal pairing of left and right eigenfunctions, and its coefficients inherit the corresponding normalization. As discussed in Sec.~\ref{Sec_V_A}, this normalization may become ill conditioned near self-orthogonality, so the coefficient magnitude need not track the actual radiated contribution.

We introduce a new modal definition based on the transmit--receive reciprocity principle in antenna theory. For a reciprocal system, if a mode radiates most strongly in an observation direction $\hat{\boldsymbol{r}}$, it should also maximize reception from the corresponding incident direction $\hat{\boldsymbol{k}}=-\hat{\boldsymbol{r}}$. In addition, polarization matching and phase conjugation are required to ensure maximum power transfer \cite{ref_TR1,ref_TR2,ref_TR3}. Considering all observation directions, the radiation pattern $\boldsymbol{F}_n(\hat{\boldsymbol{r}})$ of a mode therefore naturally determines the optimally matched incident-wave amplitude distribution for reception:
\begin{equation}
(\Theta \boldsymbol{F}_n)(\hat{\boldsymbol{r}})
=
\boldsymbol{F}_n^{*}(-\hat{\boldsymbol{r}}).
\label{eq:time_reversal_operator}
\end{equation}
Under the assumed time-harmonic convention, the operator $\Theta$ admits the physical interpretation of time reversal \cite{ref_QuantumPhysics1,ref_QuantumPhysics2}. We therefore define time-reversal characteristic modes as far-field patterns $\boldsymbol{F}_n$ that satisfy
\begin{equation}
\mathcal{S}\Theta \boldsymbol{F}_n
=
\sigma_n \boldsymbol{F}_n .
\label{eq:trcm_operator}
\end{equation}
The corresponding integral form is
\begin{equation}
\int \mathrm{d}\hat{\boldsymbol{k}}\,
\bar{\boldsymbol{S}}(\hat{\boldsymbol{r}},\hat{\boldsymbol{k}})
\cdot
\boldsymbol{F}_n^{*}(-\hat{\boldsymbol{k}})
=
\sigma_n \boldsymbol{F}_n(\hat{\boldsymbol{r}}).
\label{eq:trcm_integral}
\end{equation}
Since $\Theta$ is an antilinear operator, \eqref{eq:trcm_operator} is an antilinear eigenvalue problem. By properly choosing the global phase of $\boldsymbol{F}_n$, the eigenvalue $\sigma_n$ can be made real and nonnegative.

\begin{figure*}[!t]
  \centering
  \includegraphics[]{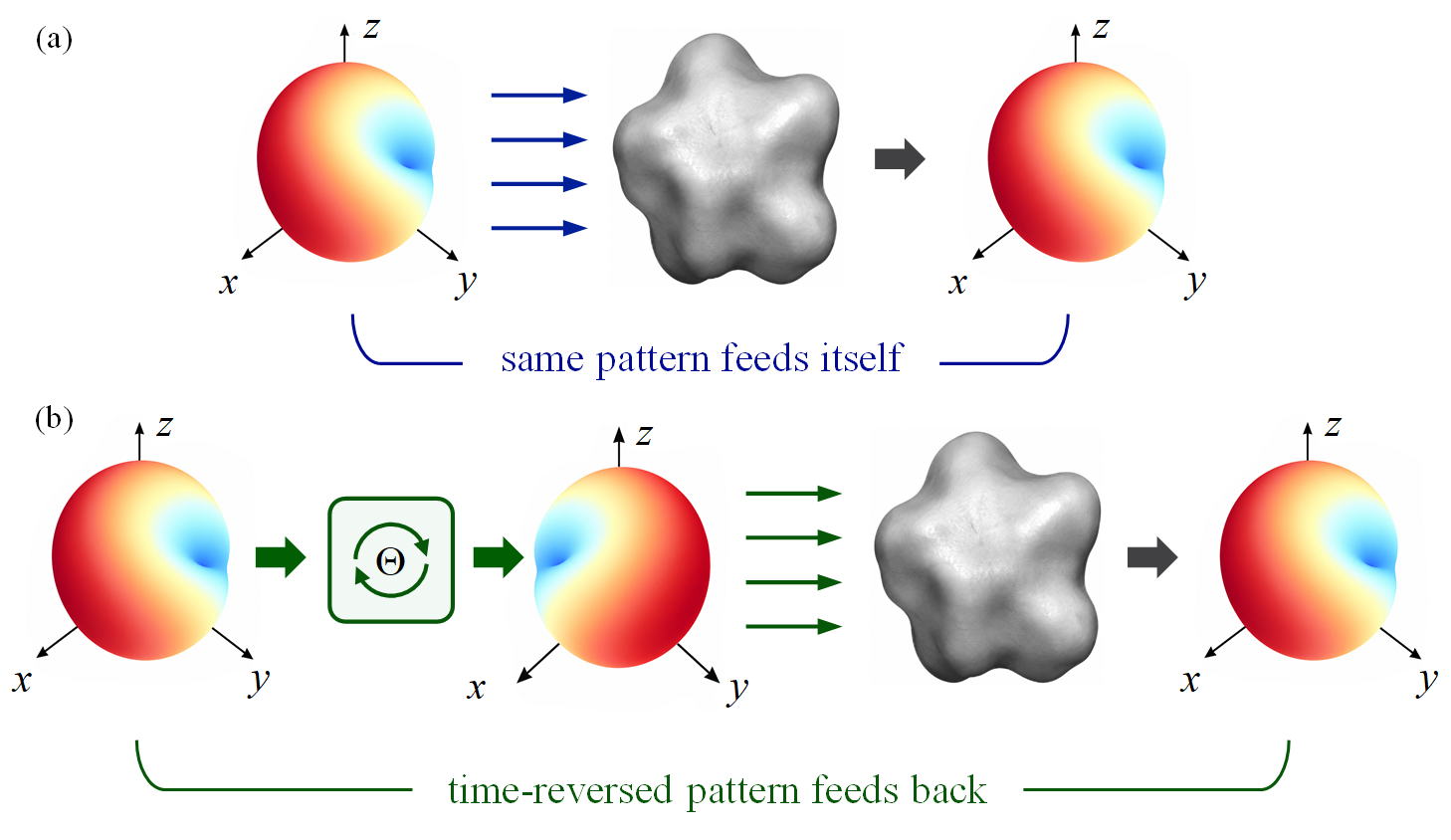}
  \caption{Comparison between classical characteristic modes and time-reversal characteristic modes. 
  (a) Classical CMT seeks a pattern that reproduces itself after scattering. 
  (b) Time-reversal CMT seeks a radiated pattern whose time-reversed counterpart, when fed back to illuminate the structure, regenerates the original pattern.}
  \label{f_pic_modes}
\end{figure*}

Fig.~\ref{f_pic_modes} compares the physical pictures of time-reversal characteristic modes and classical characteristic modes. The classical equation \eqref{eq:classical_cm_operator} seeks an incident complex-amplitude distribution that is reproduced by scattering, up to a complex scalar factor $t_n$. In contrast, the time-reversal characteristic-mode equation seeks a radiated far field $\boldsymbol{F}_n$ such that, when its time-reversed counterpart $\Theta\boldsymbol{F}_n$ is sent back to illuminate the structure, the original far field $\boldsymbol{F}_n$ is regenerated, up to a scalar factor $\sigma_n$.

\section{Properties of the Time-Reversal Characteristic-Mode Equation}

For algebraic convenience, the far-field patterns are expanded in a set of normalized vector spherical wave functions \cite{ref_Sca}. Let $\mathbf{a}$ denote the expansion coefficients of the incident wave, and let $\mathbf{f}$ denote those of the scattered wave. They are related by
\begin{equation}
\mathbf{f}=\mathbf{T}\mathbf{a}
\label{eq:T_matrix_relation}
\end{equation}
where $\mathbf{T}$ is the transition matrix \cite{ref_Unified_I,ref_Sca,ref_Waterman}. For reciprocal structures, \eqref{eq:reciprocity_sdyadic} is equivalent to the complex symmetry of $\mathbf{T}$, namely
\begin{equation}
\mathbf{T}^{\mathrm{T}}=\mathbf{T}.
\label{eq:T_complex_symmetric}
\end{equation}
Equation~\eqref{eq:trcm_operator} can then be equivalently transformed into the following algebraic matrix form:
\begin{equation}
\mathbf{T}\mathbf{f}_{n}^{*}
=
-\sigma_n \mathbf{f}_n.
\label{eq:trcm_T_matrix}
\end{equation}
The minus sign originates from the antisymmetry of the far-field asymptotic form of vector spherical harmonics under time reversal, \cf Appendix~\ref{App_A}.

\subsection{Radiation Orthogonality}

Taking the complex conjugate of both sides of \eqref{eq:trcm_T_matrix} and using \eqref{eq:T_complex_symmetric}, we obtain
\begin{equation}
\mathbf{T}^{\mathrm{H}}\mathbf{f}_n
=
-\sigma_n \mathbf{f}_{n}^{*}.
\end{equation}
Further left-multiplying by $\mathbf{T}$ gives
\begin{equation}
\mathbf{T}\mathbf{T}^{\mathrm{H}}\mathbf{f}_n
=
\sigma_n^2 \mathbf{f}_n .
\label{eq:TTH_eigenproblem}
\end{equation}
Since $\mathbf{T}\mathbf{T}^{\mathrm{H}}$ is a Hermitian positive-semidefinite matrix, its eigenvectors $\mathbf{f}_n$ are mutually orthogonal. Therefore, they can be normalized as
\begin{equation}
\mathbf{f}_{m}^{\mathrm{H}}\mathbf{f}_n
=
\delta_{mn}.
\label{eq:discrete_orthogonality}
\end{equation}
This orthogonality corresponds to radiation-power orthogonality in the continuous far-field representation:
\begin{equation}
\frac{1}{Z_0}
\int \mathrm{d}\hat{\boldsymbol{k}}\,
\boldsymbol{F}_{m}^{*}(\hat{\boldsymbol{k}})
\cdot
\boldsymbol{F}_n(\hat{\boldsymbol{k}})
=
\delta_{mn}
\label{eq:continuous_radiation_orthogonality}
\end{equation}
where $Z_0$ is the free-space wave impedance. It is worth emphasizing that this orthogonality relies only on reciprocity.

\subsection{Eigenvalue Disk and Resonance Characterization in Passive Systems}

If the scatterer is passive, energy conservation imposes the following constraint on $\mathbf{T}$ \cite[Ch.~7]{ref_Sca}:
\begin{equation}
\mathbf{T}^{\mathrm{H}}\mathbf{T}
+
\Re\{\mathbf{T}\}
\preceq
0 .
\label{eq:passivity_constraint}
\end{equation}
This inequality implies that, for any time-reversal characteristic mode $\mathbf{f}_n$, its scattering strength $\sigma_n$, or modal significance, is constrained by passivity as
\begin{equation}
\sigma_n \in [0,1]
\end{equation}
as shown in Appendix~\ref{App_B}.

\begin{figure}[!t]
  \centering
  \includegraphics[]{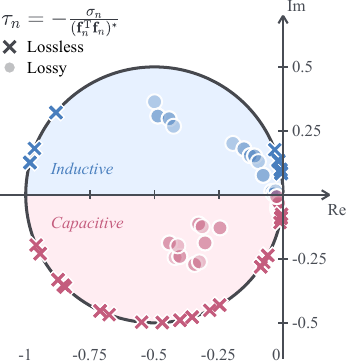}
  \caption{Location of $\tau_n$ in the complex plane computed from a two-sphere system. Lossless case with $\varepsilon_1=8.7$ and $\varepsilon_2=17.22$; lossy case with $\varepsilon_1=8.7-2.75\mathrm{j}$ and $\varepsilon_2=17.22-4.5\mathrm{j}$. The two spheres have radius $R=1012.4~\mathrm{mm}$, and their center-to-center distance is $3310.3~\mathrm{mm}$. }
  \label{fig:tau_disk}
\end{figure}

Since $\sigma_n$ is real, it does not provide the same type of modal-resonance characterization as the classical characteristic value $t_n$. To this end, we introduce the mapping
\begin{equation}
\tau_n
=
-\sigma_n
\frac{\mathbf{f}_{n}^{\mathrm{H}}\mathbf{f}_n}
{\left(\mathbf{f}_{n}^{\mathrm{T}}\mathbf{f}_n\right)^{*}}
=
-\frac{\sigma_n}
{\left(\mathbf{f}_{n}^{\mathrm{T}}\mathbf{f}_n\right)^{*}}
\label{eq:tau_definition}
\end{equation}
where the second equality follows from the normalization in \eqref{eq:discrete_orthogonality}. In Appendix~\ref{App_B}, we prove that $\tau_n$ satisfies
\begin{equation}
\left|\tau_n+\frac{1}{2}\right|
\le
\frac{1}{2}.
\label{eq:tau_disk}
\end{equation}
That is, $\tau_n$ is confined within a disk centered at $-1/2$ on the real axis with radius $1/2$. In the lossless case, equality in \eqref{eq:tau_disk} is attained, and $\tau_n$ lies exactly on the boundary of the disk, as shown in Fig.~\ref{fig:tau_disk}.

The physical interpretation of $\tau_n$ is consistent with that of the classical eigenvalue $t_n$. For purely electric materials, including PECs, the cycle-mean complex power of mode $n$ is
\begin{equation}
P_{c,n}
=
-\frac{1}{2}\mathbf{f}_{n}^{\mathrm{H}}\mathbf{a}_n
=
\frac{1}{2\sigma_n}
\left(\mathbf{f}_{n}^{\mathrm{T}}\mathbf{f}_n\right)^{*}
=
-\tau_n^{-1}P_{\mathrm{rad},n}
\label{eq:complex_power_tau}
\end{equation}
where
\begin{equation}
P_{\mathrm{rad},n}
=
\frac{1}{2}\mathbf{f}_{n}^{\mathrm{H}}\mathbf{f}_n .
\end{equation}
A resonant mode satisfies $\Im\{\tau_n\}=0$, an inductive mode satisfies $\Im\{\tau_n\}>0$, and a capacitive mode satisfies $\Im\{\tau_n\}<0$.\footnote{
This inductive/capacitive interpretation is stated for purely electric materials, including PECs. For purely magnetic materials, the correspondence is reversed; for magnetodielectric materials involving both permittivity and permeability contrasts, there is generally no simple one-to-one correspondence. The derivation can be made from the volume polarization and magnetization currents using the complex Poynting theorem; see also the discussion in \cite[Footnote 8 and Appendix A]{ref_Unified_I}.}
Therefore, the upper half of the disk in Fig.~\ref{fig:tau_disk} represents the inductive region, whereas the lower half represents the capacitive region. Note the similarity between Fig.~\ref{fig:tau_disk} and its classical counterpart in~\cite{ref_Unified_I}. However, the quantity $\tau_n$ introduced here is not the classical eigenvalue $t_n$, except in the lossless case, where the definition in \eqref{eq:tau_definition} automatically reduces to $t_n$, as discussed in the following section.

\section{Connection to Classical Characteristic Modes and Inagaki (SVD) Modes}

\subsection{Connection to Classical Characteristic Modes}

We first show that the proposed definition contains classical CMT as its lossless limit. Let $\tilde{\mathbf{f}}_n$ denote a classical characteristic-mode vector, where the tilde is used to distinguish it from a time-reversal characteristic mode. For a lossless reciprocal scatterer, $\tilde{\mathbf{f}}_n$ and its complex conjugate $\tilde{\mathbf{f}}_n^{*}$ represent the same modal degree of freedom up to a phase choice \cite[Appendix B]{ref_Unified_I}. This phase freedom can be used to align the time-reversed pattern with the original modal pattern. Writing the classical eigenvalue as $t_n = |t_n| e^{\mathrm{j}\alpha_n}$, we choose the global phase of $\tilde{\mathbf{f}}_n$ as $e^{\mathrm{j}(\alpha_n+\pi)/2}$, so that
\begin{equation}
\tilde{\mathbf{f}}_n^{*}
=
- e^{-\mathrm{j}\alpha_n}\tilde{\mathbf{f}}_n .
\label{eq:classical_phase_choice}
\end{equation}
Applying the classical eigenvalue equation to this phase-fixed vector gives
\begin{equation}
\mathbf{T}\tilde{\mathbf{f}}_n^{*}
=
- e^{-\mathrm{j}\alpha_n}\mathbf{T}\tilde{\mathbf{f}}_n
=
- t_n e^{-\mathrm{j}\alpha_n}\tilde{\mathbf{f}}_n .
\label{eq:classical_to_trcm}
\end{equation}
The last expression has exactly the form of the time-reversal characteristic-mode equation \eqref{eq:trcm_T_matrix}. Hence every classical characteristic mode of a lossless reciprocal scatterer is also a time-reversal characteristic mode, with
\begin{equation}
\sigma_n = t_n e^{-\mathrm{j}\alpha_n}=|t_n| .
\end{equation}

The converse follows from the normality induced by losslessness. Together with the complex symmetry in \eqref{eq:T_complex_symmetric}, losslessness implies that $\mathbf{T}$ and $\mathbf{T}^{\mathrm{H}}$ commute. Hence,
\begin{equation*}
\mathbf{T}\cdot \mathbf{T}\mathbf{T}^{\mathrm{H}}
=
\mathbf{T}\cdot \mathbf{T}^{\mathrm{H}}\mathbf{T}
=
\mathbf{T}\mathbf{T}^{\mathrm{H}}\cdot \mathbf{T}.
\label{eq:T_commutes_TTH}
\end{equation*}
Thus, $\mathbf{T}$ also commutes with $\mathbf{T}\mathbf{T}^{\mathrm{H}}$. The eigenspaces of $\mathbf{T}\mathbf{T}^{\mathrm{H}}$ are therefore invariant under $\mathbf{T}$, and the two operators can be diagonalized in a common modal basis \cite{ref_Matrix_Analysis}. Since a time-reversal characteristic mode satisfies \eqref{eq:TTH_eigenproblem}, it can be chosen from this common basis and therefore also satisfies
\begin{equation}
\mathbf{T}\mathbf{f}_n
=
t_n\mathbf{f}_n
\label{eq:standard_T_eigenproblem}
\end{equation}
which is precisely the scattering-based definition of a classical characteristic mode.

It remains only to identify how the two eigenvalue parametrizations are related. Using the normalization $\mathbf{f}_n^{\mathrm{H}}\mathbf{f}_n=1$, the classical eigenvalue can be written as
\begin{equation*}
t_n
=
\mathbf{f}_n^{\mathrm{H}}\mathbf{T}\mathbf{f}_n
=
-\sigma_n \mathbf{f}_n^{\mathrm{T}}\mathbf{f}_n
\label{eq:t_sigma_relation}
\end{equation*}
where the second equality follows from
\begin{equation*}
\mathbf{f}_n^{\mathrm{H}}\mathbf{T}\mathbf{f}_n
=
\left(\mathbf{f}_n^{\mathrm{H}}\mathbf{T}\mathbf{f}_n\right)^{\mathrm{T}}
=
\mathbf{f}_n^{\mathrm{T}}\mathbf{T}\mathbf{f}_n^{*}.
\end{equation*}
Together with $\sigma_n=|t_n|$, this identity gives $\left|\mathbf{f}_n^{\mathrm{T}}\mathbf{f}_n\right|=1$. Therefore, the classical eigenvalue may equivalently be written as
\begin{equation}
t_n
=
-\frac{\sigma_n}
{\left(\mathbf{f}_n^{\mathrm{T}}\mathbf{f}_n\right)^{*}}
\label{eq:t_tau_equivalence}
\end{equation}
which is exactly the definition of $\tau_n$ in \eqref{eq:tau_definition}. Thus, in the lossless case, the resonance parameter $\tau_n$ introduced for time-reversal characteristic modes reduces identically to the classical characteristic eigenvalue $t_n$.

\subsection{Connection to Inagaki Modes}

Equation~\eqref{eq:TTH_eigenproblem} provides a direct bridge between time-reversal characteristic modes and the Inagaki singular-channel picture. In matrix language, \eqref{eq:TTH_eigenproblem} is the left singular-vector problem of the $\mathbf{T}$ matrix. Owing to the complex symmetry of reciprocal structures, taking the complex conjugate of \eqref{eq:TTH_eigenproblem} gives
\begin{equation}
\mathbf{T}^{\mathrm{H}}\mathbf{T}\mathbf{f}_n^{*}
=
\sigma_n^2 \mathbf{f}_n^{*}.
\label{eq:right_inagaki}
\end{equation}
Therefore, $\mathbf{f}_n^{*}$ lies in the corresponding right singular-vector subspace. The modal significance $\sigma_n$ thus has the same meaning as the Inagaki singular value, while the mode itself is selected with an additional reciprocity structure.

In this sense, the Inagaki construction identifies the singular subspaces and their channel strengths, whereas the time-reversal construction chooses the reciprocal transmit--receive pairing inside those subspaces. The defining equation \eqref{eq:trcm_T_matrix} pairs the radiated pattern $\mathbf{f}_n$ with its matched incident channel $\mathbf{f}_n^{*}$, with the sign fixed by the vector-spherical-wave convention. For a nondegenerate singular value, this condition only fixes the modal phase. In a degenerate singular subspace, however, phase choices are insufficient; a unitary transformation within the degenerate subspace must be performed to select a basis whose members are paired with their own time-reversed receiving channels.

With this reciprocal pairing, each time-reversal characteristic mode becomes a single modal object carrying three pieces of information: the outgoing radiation pattern, the matched receiving or excitation field, and the modal radiated-power contribution. This unified excitation--radiation identity is the practical advantage for characteristic-mode analysis. It allows modal selection, excitation synthesis, mode suppression, and power attribution to be performed on the same mode, and it also allows the definition to be transferred naturally from the external wave-channel space to the MoM current space.

To obtain this MoM current-space representation, we express the $\mathbf{T}$ matrix in terms of the MoM impedance operator as (\cf \cite{ref_Unified_I})
\begin{equation}
\mathbf{T}
=
-\mathbf{U}\mathbf{Z}^{-1}\mathbf{U}^{\mathrm{T}}
\label{eq:T_from_MoM}
\end{equation}
where $\mathbf{U}$ is a real projection matrix that maps the radiation field produced by $\mathbf{I}_n$ onto the corresponding spherical-wave expansion coefficients, namely
\begin{equation}
\mathbf{f}_n=-\mathbf{U}\mathbf{I}_n .
\end{equation}
Since $\mathbf{U}$ is real, one also has $\mathbf{f}_n^{*}=-\mathbf{U}\mathbf{I}_n^{*}$.

Substituting \eqref{eq:T_from_MoM} and $\mathbf{f}_n^{*}=-\mathbf{U}\mathbf{I}_n^{*}$ into \eqref{eq:trcm_T_matrix}, left-multiplying by $\mathbf{U}^{\mathrm{T}}$, and canceling the common term
\begin{equation}
\mathbf{R}_0=\mathbf{U}^{\mathrm{T}}\mathbf{U}
\end{equation}
which is positive definite and associated with radiated power, yields
\begin{equation}
\mathbf{Z}\mathbf{I}_n
=
\sigma_n^{-1}\mathbf{R}_0\mathbf{I}_n^{*}.
\label{eq:MoM_trcm}
\end{equation}
\eqref{eq:MoM_trcm} is strongly isomorphic to the classical MoM-CMT equation. The key difference in time-reversal characteristic modes is the appearance of the conjugated current $\mathbf{I}_n^{*}$ on the right-hand side, which is precisely the manifestation of the time-reversed incident condition in the MoM current space. In a pure MoM framework, $\mathbf{R}_0$ need not be formed through an explicit spherical-wave projection; it may be chosen as the radiation part associated with the impedance matrix, following the construction for the corresponding structure or integral equation in~\cite[Appendix C]{ref_Unified_I}. The characteristic currents can then be normalized as
\begin{equation}
\mathbf{I}_m^{\mathrm{H}}\mathbf{R}_0\mathbf{I}_n
=
\delta_{mn}.
\label{eq:current_radiation_normalization}
\end{equation}

Equations~\eqref{eq:trcm_integral}, \eqref{eq:trcm_T_matrix}, and \eqref{eq:MoM_trcm} constitute three commonly used equivalent definitions of time-reversal characteristic modes. Each can be used independently to solve for the modes. The definition in \eqref{eq:trcm_integral} is based entirely on plane-wave scattering problems; wrappers are available to obtain the scattering dyadic \cite{ref_OA_code2} from arbitrary commercial electromagnetic solvers that support plane-wave scattering, such as HFSS, FEKO, CST, and COMSOL. It is therefore the most general approach for computing time-reversal characteristic modes. The definition in \eqref{eq:trcm_T_matrix} is suitable when the T-matrix is readily available. For example, regular structures such as spheres have closed-form analytic T-matrix representations \cite[Ch.~8]{ref_Sca}. In addition, the T-matrix of multibody systems can be constructed semianalytically from substructures using Wigner translation and rotation formulas \cite{ref_mysyn_CMA}, making it highly efficient for such problems. The definition in \eqref{eq:MoM_trcm} can be solved entirely within the MoM framework, and a solution algorithm is provided in Appendix~\ref{App_C}. For large-scale problems, matrix-free methods such as MLFMM combined with implicitly restarted Arnoldi methods may be used to accelerate the solution of time-reversal characteristic modes, \cf \cite{ref_CM_MLFMM}.

\section{Modal Expansion Analysis}

The field expansion in terms of time-reversal characteristic modes takes a particularly simple form. The excitation $\boldsymbol{A}(\hat{\boldsymbol{k}})$ can be expanded in the characteristic excitation fields $\{\sigma_n^{-1}\Theta\boldsymbol{F}_n\}$ as
\begin{equation}
\boldsymbol{A}(\hat{\boldsymbol{k}})
=
\sum_n
w_n \sigma_n^{-1}
\boldsymbol{F}_n^{*}(-\hat{\boldsymbol{k}}).
\label{eq:incident_expansion_TRCM}
\end{equation}
Using the orthogonality relation in \eqref{eq:continuous_radiation_orthogonality}, one obtains
\begin{equation}
w_n
=
\frac{\sigma_n}{Z_0}
\int \mathrm{d}\hat{\boldsymbol{k}}\,
\boldsymbol{F}_n(-\hat{\boldsymbol{k}})
\cdot
\boldsymbol{A}(\hat{\boldsymbol{k}})
=
\frac{\sigma_n}{Z_0}
\int \mathrm{d}\hat{\boldsymbol{k}}\,
\boldsymbol{F}_n(\hat{\boldsymbol{k}})
\cdot
\boldsymbol{A}(-\hat{\boldsymbol{k}}).
\label{eq:TRCM_weight_continuous}
\end{equation}
A special case is a plane wave propagating along the direction $\hat{\boldsymbol{r}}$ with amplitude $\boldsymbol{E}_0$, for which
\begin{equation}
\boldsymbol{A}(\hat{\boldsymbol{k}})
=
\boldsymbol{E}_0 \delta(\hat{\boldsymbol{r}}-\hat{\boldsymbol{k}}).
\end{equation}
The corresponding modal coefficient becomes
\begin{equation}
w_n
=
\frac{\sigma_n}{Z_0}
\boldsymbol{E}_0
\cdot
\boldsymbol{F}_n(-\hat{\boldsymbol{r}}).
\label{eq:plane_wave_weight_TRCM}
\end{equation}
Thus, under plane-wave excitation, the modal expansion coefficient is determined only by the modal field in the opposite propagation direction, \ie the receiving direction. If a modal null is aligned with the direction of arrival, no reception is produced.

Modal coefficients can also be obtained in the T-matrix formulation, where the excitation $\boldsymbol{A}$ is expanded into a coefficient vector of regular spherical waves, denoted by $\mathbf{a}$. Equation~\eqref{eq:TRCM_weight_continuous} is then equivalently written as
\begin{equation}
w_n
=
\sigma_n \mathbf{f}_n^{\mathrm{T}}\mathbf{a}.
\label{eq:TRCM_weight_Tmatrix}
\end{equation}

When the excitation is local to the structure, as is common in antenna engineering, for example in voltage-port feeding, the MoM-based definition in \eqref{eq:MoM_trcm} should be used to determine the expansion coefficients. Let the local excitation be represented by the MoM excitation vector $\mathbf{V}$, and let the induced current be expanded as
\begin{equation}
\mathbf{I}
=
\sum_n w_n \mathbf{I}_n .
\end{equation}
Using $\mathbf{Z}\mathbf{I}=\mathbf{V}$ together with \eqref{eq:MoM_trcm} and \eqref{eq:current_radiation_normalization}, one obtains
\begin{equation}
w_n
=
\sigma_n \mathbf{I}_n^{\mathrm{T}}\mathbf{V}.
\label{eq:TRCM_weight_MoM}
\end{equation}

Owing to the radiation-power orthogonality of time-reversal characteristic modes, the following Parseval identity holds:
\begin{equation}
P_{\mathrm{rad}}
=
\frac{1}{2}\sum_n |w_n|^2 .
\label{eq:Parseval_TRCM}
\end{equation}
Each term $\tfrac{1}{2}|w_n|^2$ measures the total radiated power contributed by that mode. Modes with larger $|w_n|$ are therefore dominant contributors. This property is important for modal engineering: for a desired mode, one may maximize $|w_n|$, whereas, for a mode to be suppressed, one may minimize it. This interpretation is fully consistent with the intuition underlying typical mode-based design \cite{ref_CM_APP1,ref_CM_APP2,ref_CM_APP3,ref_CM_APP4}.

However, this modal-weight intuition fails for classical characteristic modes in lossy systems. Once radiation-power orthogonality is lost, a mode with a large classical expansion coefficient may contribute only weakly to the radiated power. Interpreting such a coefficient as modal dominance is therefore misleading.

\subsection{Issue of Classical Characteristic-Mode Expansion in Lossy Systems}
\label{Sec_V_A}

In dissipative systems, modal expansions must rely on biorthogonality, \ie the orthogonality between left and right eigenvectors. The resulting coefficient normalization can be seen clearly in the T-matrix framework. The left eigenvectors $\tilde{\mathbf{g}}_n$ and right eigenvectors satisfy the biorthogonality relation \cite{ref_Matrix_Analysis}
\begin{equation}
\tilde{\mathbf{g}}_m^{\mathrm{H}}
\tilde{\mathbf{f}}_n
=
c_0\delta_{mn}.
\end{equation}
Since $\mathbf{T}$ is symmetric, the left eigenvector can be written as
\begin{equation}
\tilde{\mathbf{g}}_n
=
\tilde{\mathbf{f}}_n^{*}.
\end{equation}
Therefore, the biorthogonality relation can be reduced to the following unconjugated inner product \cite[eq.~(26)]{ref_Unified_II}:
\begin{equation}
\tilde{\mathbf{f}}_m^{\mathrm{T}}
\tilde{\mathbf{f}}_n
=
c_0\delta_{mn}
\label{eq:classical_biorthogonality}
\end{equation}
where $c_0$ is a complex normalization scalar. The corresponding expansion coefficient can then be written as \cite[eq.~(44)]{ref_Unified_I}
\begin{equation}
w_n
=
t_n
\frac{
\tilde{\mathbf{f}}_n^{\mathrm{T}}\mathbf{a}
}{
\tilde{\mathbf{f}}_n^{\mathrm{T}}\tilde{\mathbf{f}}_n
}.
\label{eq:classical_weight_lossy}
\end{equation}
The validity of \eqref{eq:classical_weight_lossy} requires self-orthogonality to be avoided, namely
\begin{equation}
\tilde{\mathbf{f}}_n^{\mathrm{T}}\tilde{\mathbf{f}}_n
=
c_0
\neq
0.
\end{equation}
Because loss renders the $\mathbf{T}$ matrix nonnormal, certain parameter values, such as material parameters, geometrical parameters, or frequencies, may make $\mathbf{T}$ defective. Such points are known as exceptional points. Defective modes are self-orthogonal, which causes the denominator in \eqref{eq:classical_weight_lossy} to vanish and hence makes the expansion coefficient $w_n$ diverge.

\subsection{Case 1: Two Lossy Spheres}

\begin{figure}[!t]
  \centering
  \includegraphics[]{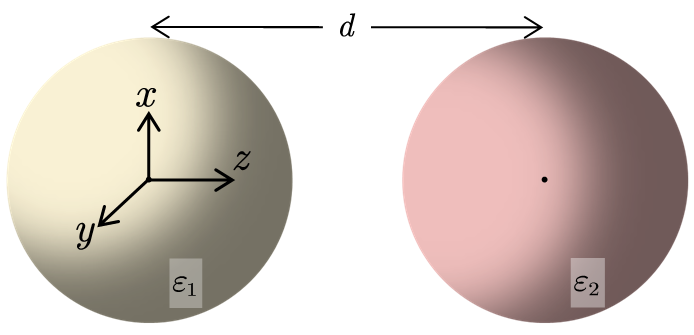}
  \caption{Geometry and parameters of the two-sphere system. Each dielectric sphere has radius $R=1012.4~\mathrm{mm}$; the relative permittivities are $\varepsilon_1=8.7-2.75\mathrm{j}$ and $\varepsilon_2=17.22-4.5\mathrm{j}$; and the center-to-center separation is $d=3310.3~\mathrm{mm}$.}
  \label{fig:two_spheres}
\end{figure}

\begin{figure}[!t]
  \centering
  \includegraphics[]{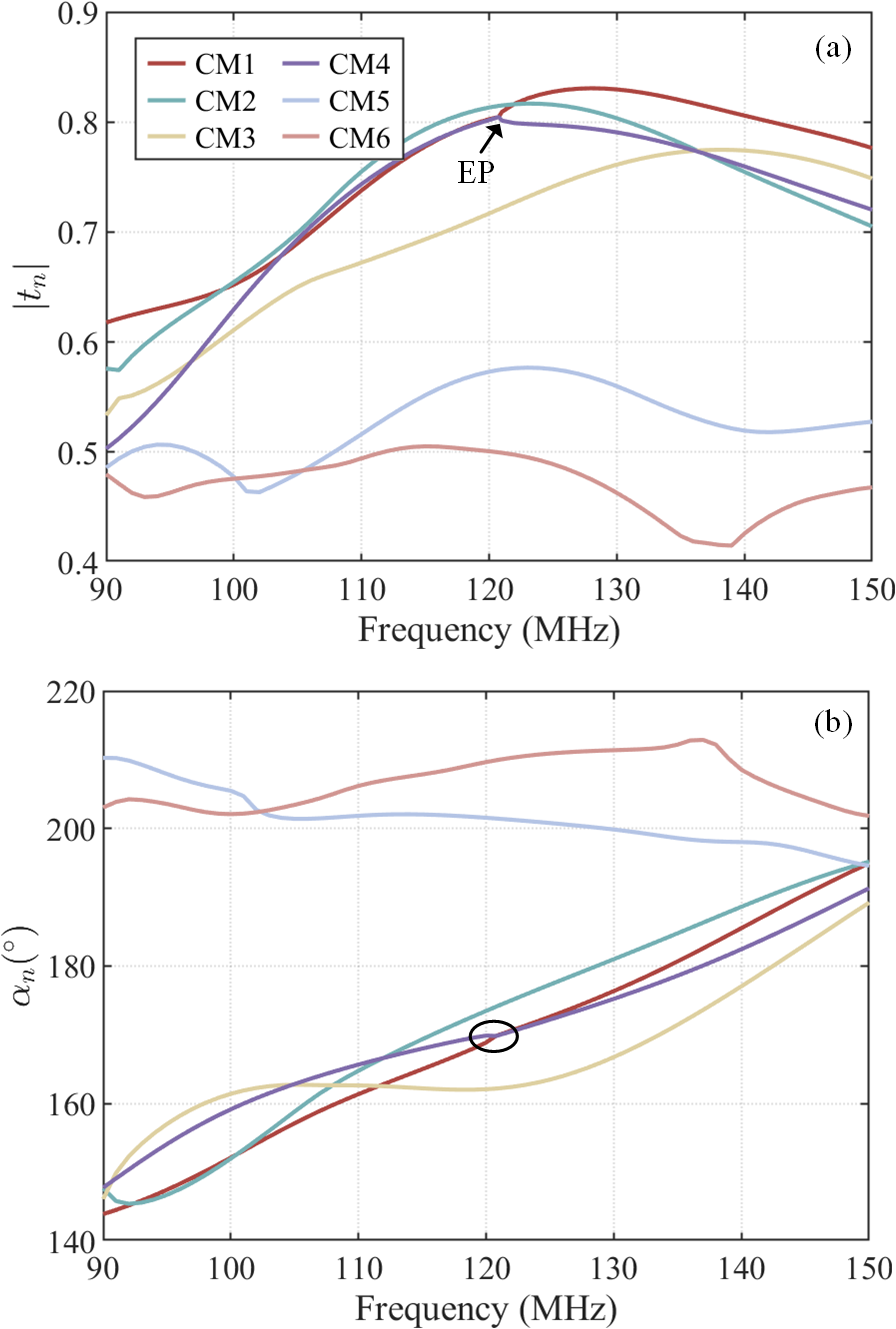}
    \caption{Eigenvalue trajectories of the first six classical characteristic modes for the lossy two-sphere system. 
  (a) Magnitudes and (b) phases of $t_n$. The EP is marked near 120~MHz.}
  \label{fig:f_bisphMS}
\end{figure}

We first use the two lossy dielectric spheres in Fig.~\ref{fig:two_spheres} to demonstrate the issue of classical characteristic-mode expansion. The $\mathbf{T}$ matrix of this system is obtained semianalytically from the Mie-series solution combined with Wigner translation formulas \cite{ref_mysyn_CMA,ref_myBCM}. In this calculation, the minimum degree of spherical-wave expansion is 17.

\begin{figure}[!t]
  \centering
  \includegraphics[]{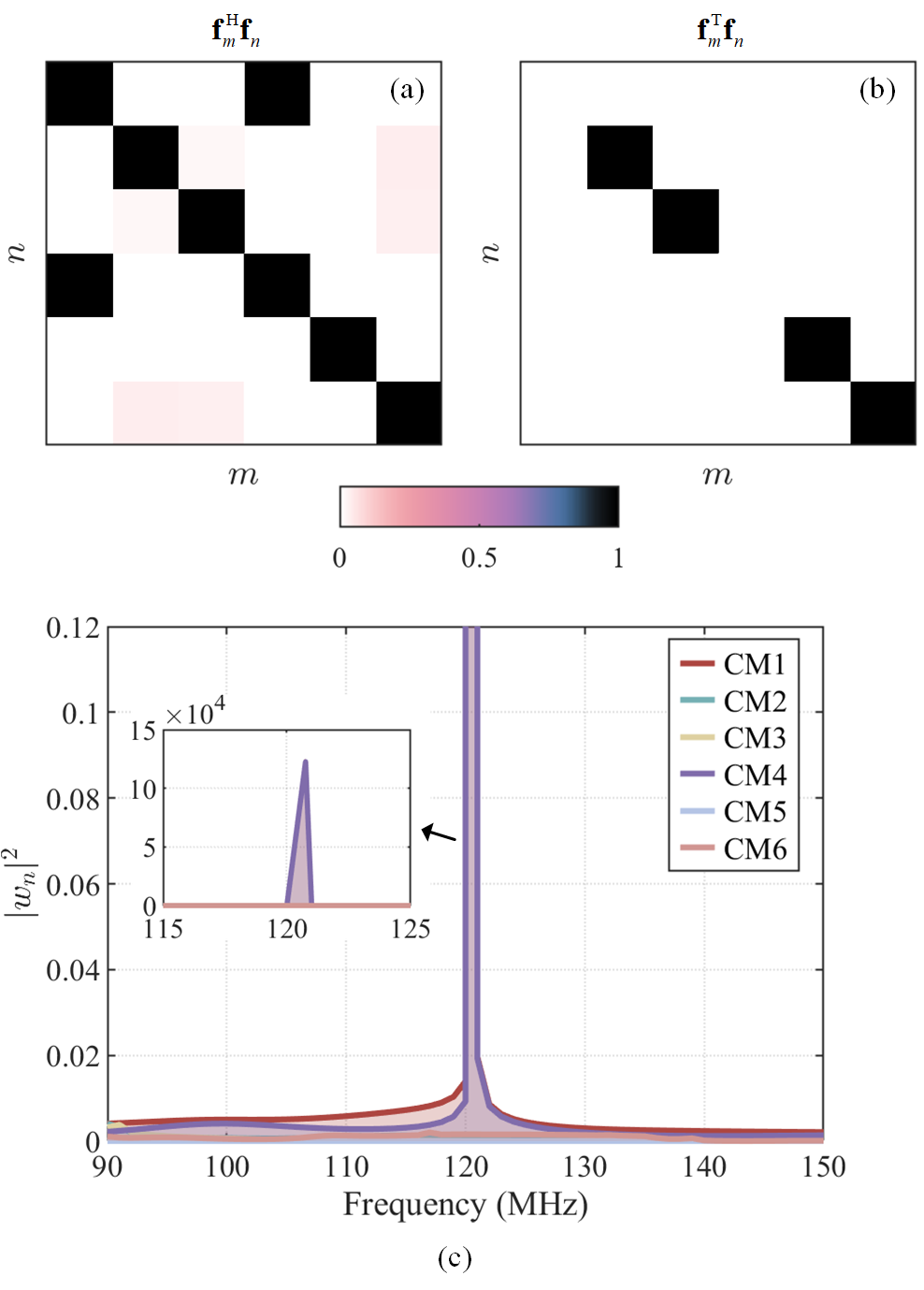 }
    \caption{Breakdown of the classical characteristic-mode expansion near the EP of the lossy two-sphere system. 
  (a) Hilbert inner products $|\mathbf{f}_m^{\mathrm{H}}\mathbf{f}_n|$ of the first six classical characteristic modes, showing the coalescence of modes 1 and 4. 
  (b) Reaction products $|\mathbf{f}_m^{\mathrm{T}}\mathbf{f}_n|$, where the vanishing diagonal terms of the coalesced modes indicate self-orthogonality. 
  (c) Modal expansion coefficients $|w_n|^2$ under the incident plane wave excitation, which become singular near the EP.}
  \label{fig:classical_MWC}
\end{figure}

In the frequency range from 90 to 150~MHz, eigentraces of the first six characteristic modes are calculated, as shown in Fig.~\ref{fig:f_bisphMS}. The exceptional point is marked as ``EP'' and occurs at approximately 120~MHz, more precisely 120755660~Hz, where the eigenvalues $t_n$ of modes 1 and 4 coalesce. We further compute the Hilbert inner product between modes, $\rho_{mn}
=
\left|
\mathbf{f}_m^{\mathrm{H}}\mathbf{f}_n
\right|$, as shown in Fig.~\ref{fig:classical_MWC}(a). It is found that $\rho_{14}\approx 1$, indicating that the eigenvectors of modes 1 and 4 also coalesce and become collinear. It should be noted that modes 1 and 4 both belong to the irreducible symmetry group with angular moment number $m=0$, and therefore there is no geometry-induced degeneracy. This can also be identified from the eigenvalue trajectories versus frequency in Fig.~\ref{fig:f_bisphMS}. Hence, the degeneracy between modes 1 and 4 near 120~MHz can be identified as a 2nd-order EP (EP2) degeneracy. Since modes at an EP are self-orthogonal, one has $\mathbf{f}_1^{\mathrm{T}}\mathbf{f}_1
=
\mathbf{f}_4^{\mathrm{T}}\mathbf{f}_4
\approx
0$, as shown in Fig.~\ref{fig:classical_MWC}(b). This leads to the divergence of the expansion coefficients determined by \eqref{eq:classical_weight_lossy}.

To clearly demonstrate the modal expansion coefficients, we consider a TE-polarized plane wave incident from the direction $\theta=45^{\circ}$. Figure~\ref{fig:classical_MWC}(c) shows the evolution of the modal expansion coefficients over the entire frequency band. In both the low- and high-frequency regions, the values of $|w_n|^2$ for all modes are below 0.02. However, in the vicinity of 120~MHz, the $|w_n|^2$ values of modes 1 and 4 rapidly blow up to the order of $10^5$.

\begin{figure}[!t]
  \centering
  \includegraphics[]{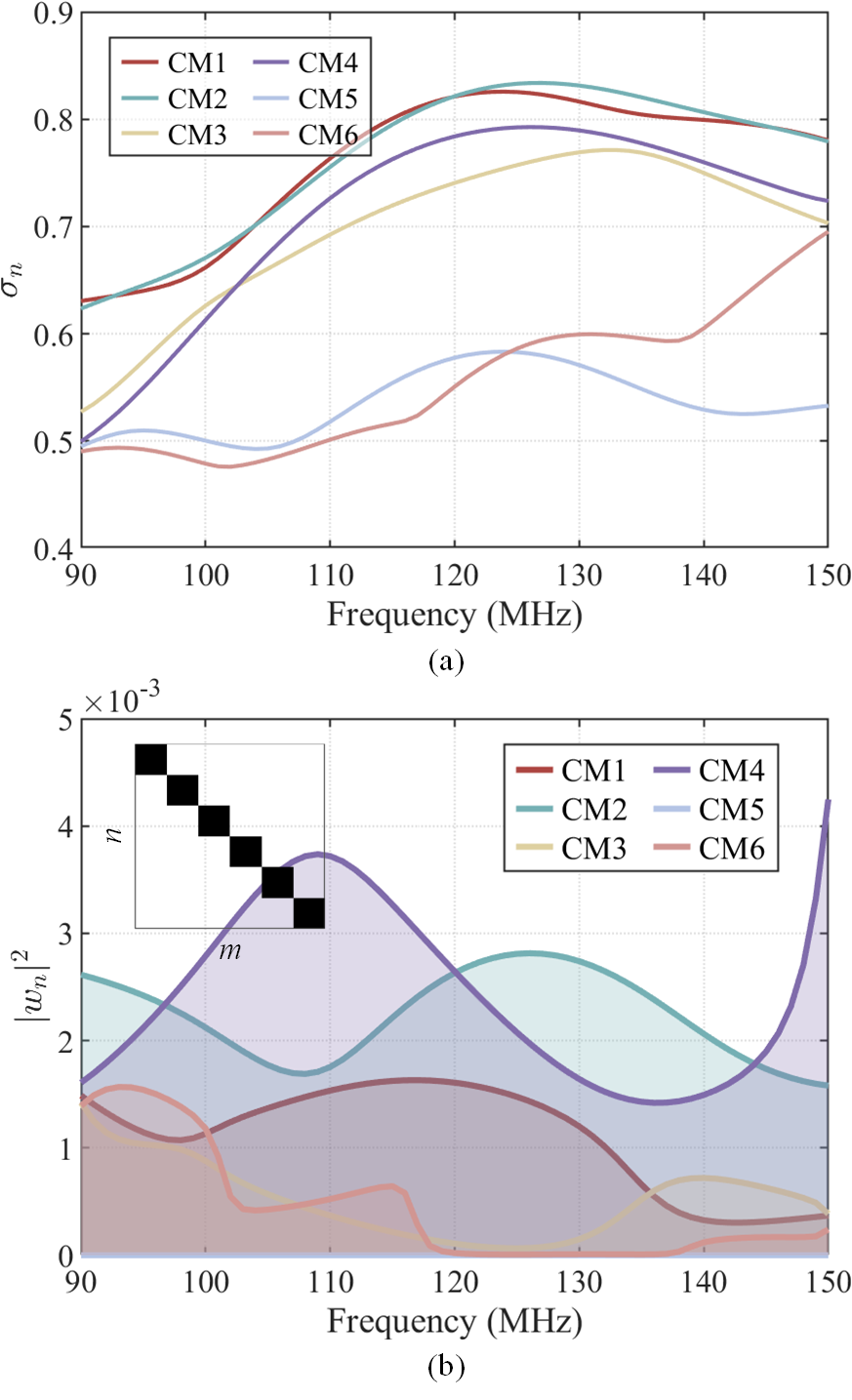} 
    \caption{Time-reversal characteristic-mode results for the lossy two-sphere system. 
  (a) Modal-significance trajectories $\sigma_n$ of the first six time-reversal characteristic modes. 
  (b) Modal expansion coefficients $|w_n|^2$ under the same incident plane-wave excitation used in Fig.~\ref{fig:classical_MWC}. 
  The inset shows the Hilbert inner products $|\mathbf{f}_m^{\mathrm{H}}\mathbf{f}_n|$, confirming the radiation-power orthogonality of the time-reversal characteristic modes.}
  \label{fig:TRCM_spheres}
\end{figure}

Figure~\ref{fig:TRCM_spheres} presents the results for the first six time-reversal characteristic modes. Panel (a) shows the trajectories of $\sigma_n$, and panel (b) shows the expansion coefficients of these six modes under the same incident excitation, together with the radiation-power orthogonality relation. Comparing Figs.~\ref{fig:f_bisphMS} and~\ref{fig:classical_MWC} with Fig.~\ref{fig:TRCM_spheres} shows that time-reversal characteristic modes remain suitable for modal decomposition even in systems containing an EP.

\subsection{Case 2: Folded Antenna}

\begin{figure}[!t]
  \centering
  \includegraphics[]{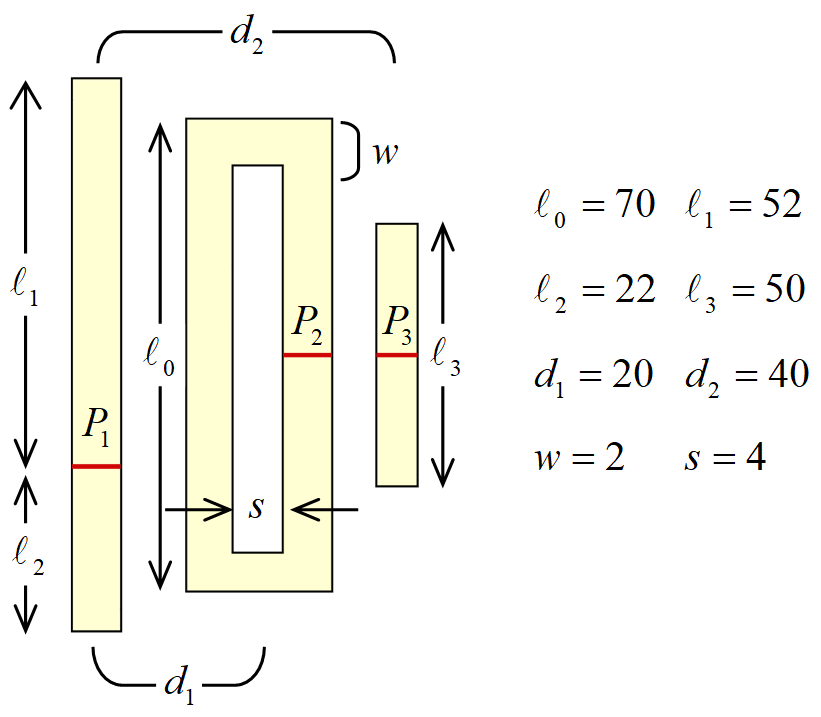} 
  \caption{Geometry and parameters of the folded antenna. The loading positions are denoted by $P_1$, $P_2$, and $P_3$. The dimensions are all in millimeters.}
  \label{fig:folded_antenna}
\end{figure}

\begin{figure}[!t]
  \centering
  \includegraphics[]{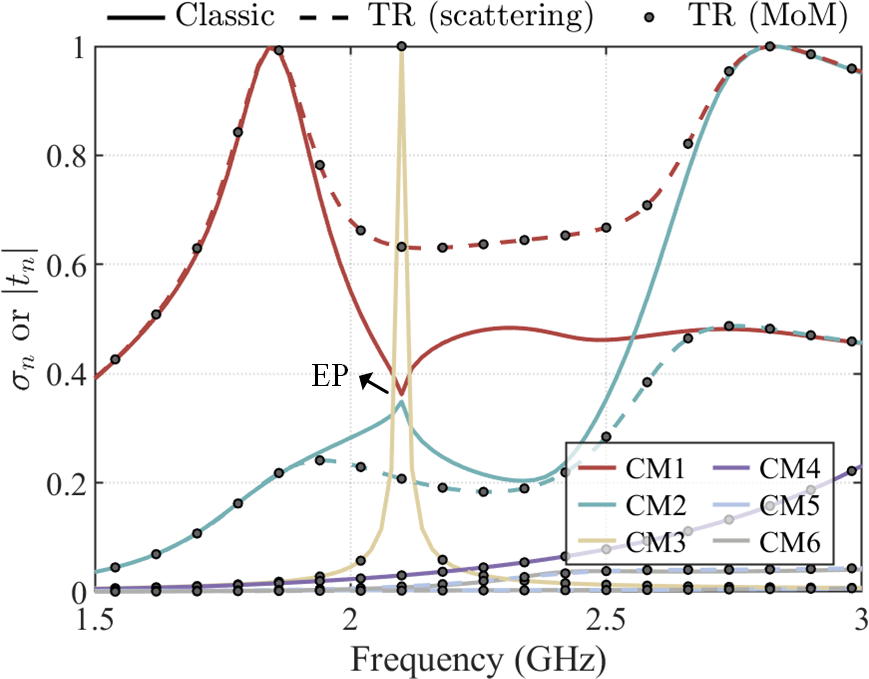} 
  \caption{Eigenvalue trajectories of the folded antenna. Solid lines denote the magnitudes $|t_n|$ of the classical characteristic eigenvalues, dashed lines denote the time-reversal modal significances $\sigma_n$ computed from the scattering-based formulation, and circular markers denote $\sigma_n$ computed from the MoM formulation. An EP of the classical characteristic modes is observed near 2.1~GHz.}
  \label{fig:antenna_MS}
\end{figure}

The next example considers the loaded folded antenna shown in Fig.~\ref{fig:folded_antenna}; its specific geometric and loading parameters are given in the figure. The antenna is loaded with a $1.6$~nH inductor at $P_1$, a $434$~$\Omega$ resistor at $P_2$, and a $14.8$~pF capacitor at $P_3$. In the frequency range from 1.5 to 3~GHz, Fig.~\ref{fig:antenna_MS} shows the eigenvalue trajectories of the classical characteristic modes and those of the time-reversal characteristic modes. The time-reversal characteristic modes are computed using both the scattering-based formulation and the MoM-based formulation via Appendix~\ref{App_C}. The MoM-based trajectories agree with those obtained from the scattering-based formulation, verifying the equivalence between the scattering-based and MoM-based formulations.

A closer inspection of Fig.~\ref{fig:antenna_MS} reveals a clear difference between the two types of modes. Classical characteristic modes 1 and 2 exhibit a tracking error near 2.1~GHz, where their eigenvalue trajectories cross. This occurs because 2.1~GHz is precisely the frequency at which modes 1 and 2 undergo an EP2 degeneracy, causing the two modes to become nearly collinear. Consequently, the mode-tracking algorithm based on the correlation coefficient loses its underlying assumption, since the modes are no longer independent \cite{ref_CM_track1,ref_CM_track2,ref_CM_track3}. In contrast, the time-reversal characteristic modes do not suffer from such a tracking error, and their eigenvalue trajectories can be smoothly tracked over the entire frequency band.

\begin{figure}[!t]
  \centering
  \includegraphics[]{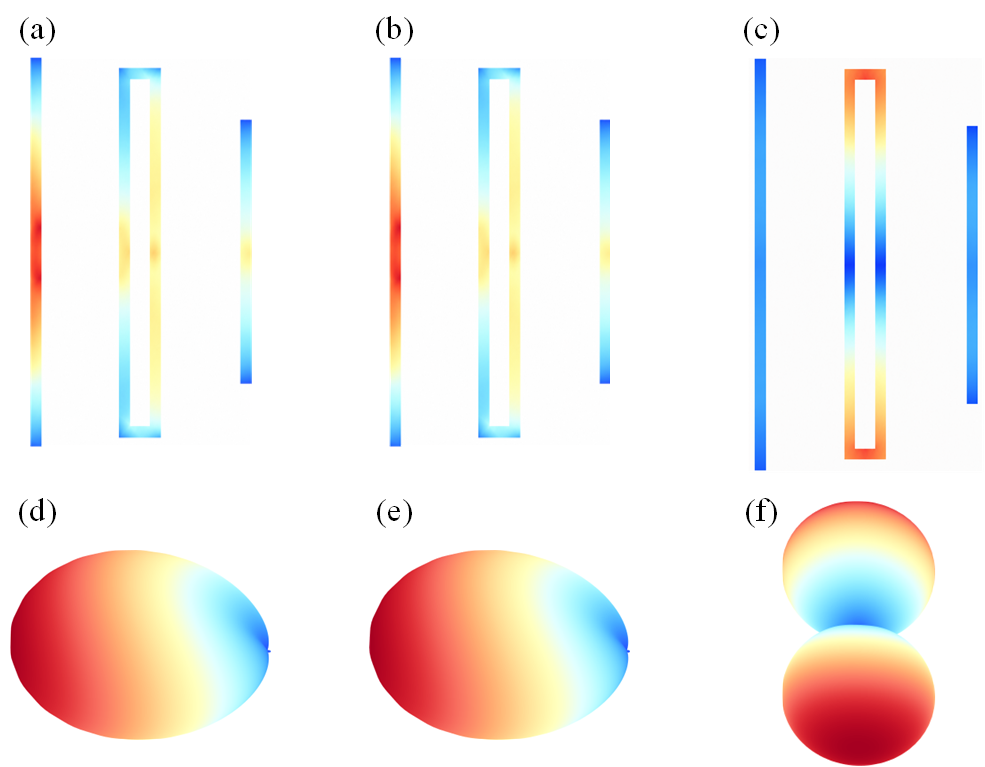} 
  \caption{Current distributions and radiation patterns of the first three classical characteristic modes at 2.1~GHz. Panels (a)--(c) show the modal current distributions, and panels (d)--(f) show the corresponding radiation patterns. Modes 1 and 2 exhibit nearly identical current distributions and radiation patterns, providing direct evidence of the EP2 degeneracy, whereas mode 3 remains distinct.}
  \label{fig:classical_currents_patterns}
\end{figure}

\begin{figure}[!t]
  \centering
  \includegraphics[]{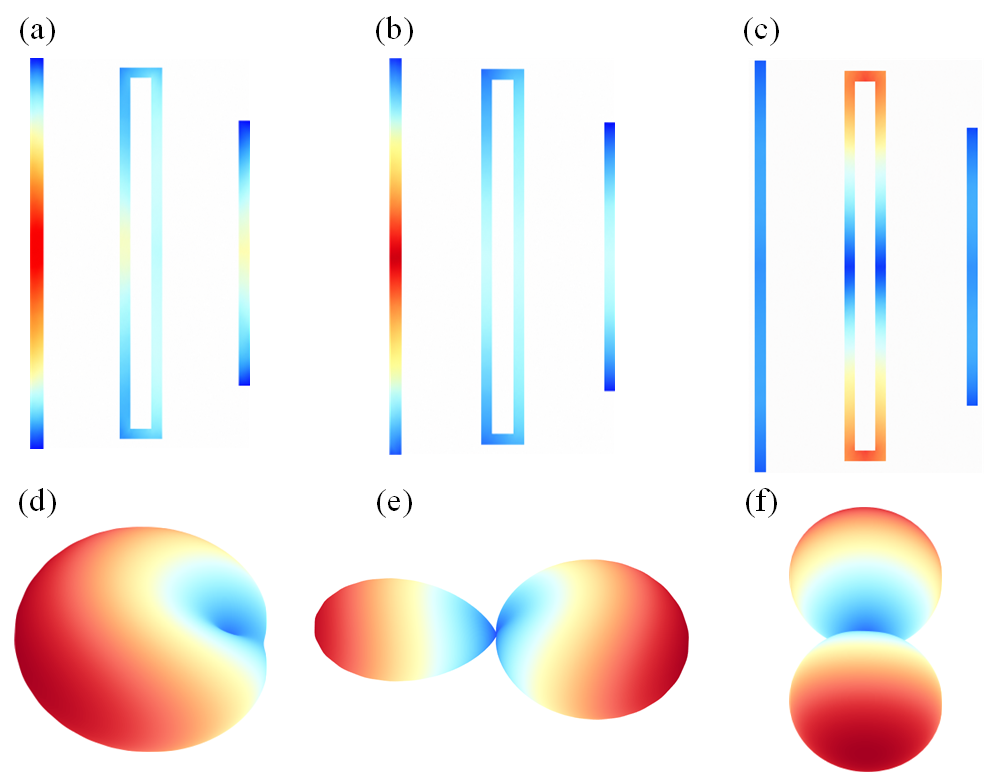} 
  \caption{Current distributions and radiation patterns of the first three time-reversal characteristic modes at 2.1~GHz. Panels (a)--(c) show the modal current distributions, and panels (d)--(f) show the corresponding radiation patterns. Unlike the classical characteristic modes in Fig.~\ref{fig:classical_currents_patterns}, the time-reversal characteristic modes remain distinct, demonstrating that the proposed modes provide a complete basis.}
  \label{fig:TRCM_currents_patterns}
\end{figure}

Figure~\ref{fig:classical_currents_patterns} shows the current distributions and radiation patterns of the first three classical characteristic modes at 2.1~GHz. Comparing panels (a) and (b) with panels (d) and (e), one finds that the current distributions and radiation patterns of the two modes are almost identical. This provides an intuitive verification of the EP2 degeneracy. By contrast, Fig.~\ref{fig:TRCM_currents_patterns} shows that the first three time-reversal characteristic modes have distinct current distributions and radiation patterns. This also indicates that time-reversal characteristic modes remain complete, whereas classical characteristic modes become incomplete at the EP due to modal collinearity and the resulting loss of one dimension.

It is worth noting that mode 3 of the time-reversal characteristic modes coincides exactly with the corresponding classical characteristic mode, both spectrally and in terms of current distribution and radiation pattern. This is because the loss in this structure is introduced by the resistor loaded at $P_2$, whereas mode 3 has a current null at this loading position and is therefore unaffected. In addition, since $\sigma_n=|t_n|$ in the lossless case, a pronounced deviation between $\sigma_n$ and $|t_n|$ indicates that the corresponding mode is affected by dissipation. For example, Fig.~\ref{fig:antenna_MS} shows that, at 2.1~GHz, modes 1 and 2 exhibit clear differences between $\sigma_n$ and $|t_n|$, since appreciable current intensities of these modes occur at $P_2$, where a dissipative element is loaded, as confirmed by Figs.~\ref{fig:classical_currents_patterns} and~\ref{fig:TRCM_currents_patterns}.

\begin{figure}[!t]
  \centering
  \includegraphics[]{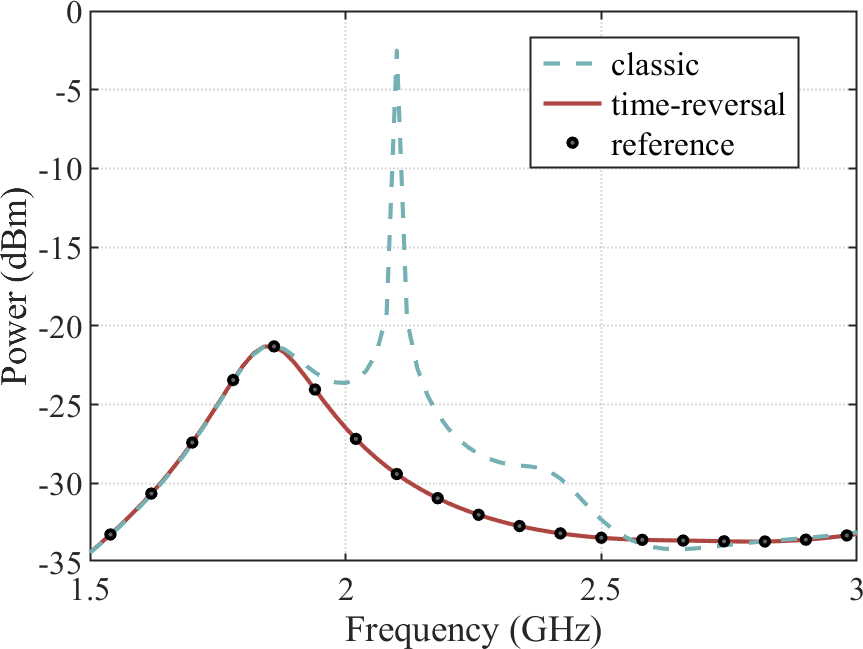}
  \caption{Radiated power of the folded antenna excited by a 1-V voltage source at $P_2$. The dashed line denotes the result obtained from the classical characteristic-mode expansion, the solid line denotes the result obtained from the time-reversal characteristic-mode expansion, and the circular markers denote the reference solution. The classical characteristic-mode result exhibits a sharp divergence near the exceptional-point region, whereas the time-reversal characteristic-mode result remains in close agreement with the reference over the entire frequency band.}
  \label{fig:antenna_power}
\end{figure}

Finally, we apply a 1-V voltage source excitation at $P_2$ and compute the sum of the radiated-power contributions of all modes, as shown in Fig.~\ref{fig:antenna_power}. In the low-frequency regions, the radiated power computed from the classical characteristic-mode expansion agrees with the reference result. However, near 2.1~GHz, it rapidly diverges and exceeds the actual value by approximately 27~dB. This divergence is caused by the exceptional point, which induces self-orthogonality and hence singular modal expansion coefficients. At the same time, the loss of modal radiation-power orthogonality destroys the power interpretation of modal weights. In contrast, the radiated power computed from the time-reversal characteristic-mode expansion agrees with the reference result over the entire frequency band, demonstrating its numerical stability.

\section{Conclusion}

A time-reversal characteristic-mode theory has been developed for reciprocal lossy electromagnetic structures. The proposed definition is based on a transmit--receive physical picture: the radiation pattern of a mode determines the time-reversed incident field that is optimally matched to couple energy back into the same mode. This leads to an antilinear characteristic-mode equation whose solutions remain radiation-power orthogonal even in the presence of material loss, lossy loading, or matched absorption. As a result, the modal expansion coefficients directly quantify the radiated-power contributions of the corresponding modes, rather than depending on a biorthogonal normalization of left and right eigenvectors.

Equivalent formulations have been derived in the scattering-operator, T-matrix, and MoM frameworks. The scattering and T-matrix formulations reveal the time-reversal picture in the external wave-channel space, whereas the MoM formulation transfers the same modal definition to current degrees of freedom and local excitation vectors. This makes the proposed theory applicable not only to plane-wave or spherical-wave scattering problems, but also to antenna problems involving port or lumped-source excitations. In the lossless limit, the proposed modes reduce to classical characteristic modes, showing that the present theory is a natural extension of classical CMT to reciprocal lossy systems.

The relationship with Inagaki modes has also been clarified. Time-reversal characteristic modes correspond to a reciprocity-constrained selection within the singular subspaces, in which each outgoing radiation pattern is paired with its time-reversed receiving channel. Once this pairing is fixed, each mode possesses a single excitation--radiation identity and admits an equivalent MoM current-space representation. This distinguishes the proposed modes from a generic input--output channel decomposition.

Two numerical examples were used to validate the theory. In a lossy two-sphere system, classical characteristic modes coalesce at an exceptional point, causing self-orthogonality and singular modal expansion coefficients. In contrast, the proposed modes remain radiation-power orthogonal and yield stable modal weights. In a loaded folded antenna, the scattering-based and MoM-based formulations give consistent eigenvalue trajectories, while the time-reversal characteristic-mode expansion accurately reproduces the reference radiated power over the entire frequency band. These results demonstrate that the proposed framework provides a stable, radiation-power-orthogonal, and power-interpretable modal decomposition for lossy and loaded electromagnetic structures. Since the definition relies on reciprocity rather than passivity, it also provides a basis for extending characteristic-mode analysis to reciprocal active or gain-assisted systems.

\begin{appendices}

\section{Derivation of the T-Matrix Definition of Time-Reversal Characteristic Modes}
\label{App_A}

The derivation of the time-reversal characteristic-mode definition in \eqref{eq:trcm_T_matrix} from the integral equation \eqref{eq:trcm_integral} follows steps similar to those in~\cite{ref_Sdyadic}. An arbitrary far-field complex amplitude $\boldsymbol{F}(\hat{\boldsymbol{r}})$ can be expanded in the complex vector-spherical-harmonic basis $\{\boldsymbol{\phi}_{\alpha}\}$, corresponding to the far-field asymptotic forms of vector spherical waves, as
\begin{equation}
\boldsymbol{F}(\hat{\boldsymbol{r}})
=
\sqrt{Z_0}
\sum_{\alpha}
f_{\alpha}
\boldsymbol{\phi}_{\alpha}(\hat{\boldsymbol{r}}).
\label{eq:appA_farfield_expansion}
\end{equation}
The orthogonality relation
\begin{equation}
\int
\mathrm{d}\hat{\boldsymbol{k}}\,
\boldsymbol{\phi}_{\alpha}^{*}(\hat{\boldsymbol{k}})
\cdot
\boldsymbol{\phi}_{\beta}(\hat{\boldsymbol{k}})
=
\delta_{\alpha\beta}
\label{eq:appA_basis_orthogonality}
\end{equation}
gives the expansion coefficient as
\begin{equation}
f_{\alpha}
=
\frac{1}{\sqrt{Z_0}}
\int
\mathrm{d}\hat{\boldsymbol{k}}\,
\boldsymbol{\phi}_{\alpha}^{*}(\hat{\boldsymbol{k}})
\cdot
\boldsymbol{F}(\hat{\boldsymbol{k}})
=
\frac{1}{\sqrt{Z_0}}
\int
\mathrm{d}\hat{\boldsymbol{k}}\,
\boldsymbol{\phi}_{\alpha}^{*}(-\hat{\boldsymbol{k}})
\cdot
\boldsymbol{F}(-\hat{\boldsymbol{k}}).
\label{eq:appA_coefficient}
\end{equation}
The complex vector-spherical-harmonic basis satisfies the time-reversal antisymmetry relation
\begin{equation}
\boldsymbol{\phi}_{\alpha}^{*}(-\hat{\boldsymbol{r}})
=
-\boldsymbol{\phi}_{\alpha}(\hat{\boldsymbol{r}}).
\label{eq:appA_time_reversal_basis}
\end{equation}
Taking the complex conjugate of \eqref{eq:appA_coefficient} and using \eqref{eq:appA_time_reversal_basis}, we obtain
\begin{equation}
f_{\alpha}^{*}
=
-\frac{1}{\sqrt{Z_0}}
\int
\mathrm{d}\hat{\boldsymbol{k}}\,
\boldsymbol{\phi}_{\alpha}^{*}(\hat{\boldsymbol{k}})
\cdot
\boldsymbol{F}^{*}(-\hat{\boldsymbol{k}}).
\label{eq:appA_conjugated_coefficient}
\end{equation}

The scattering dyadic
$\bar{\boldsymbol{S}}(\hat{\boldsymbol{r}},\hat{\boldsymbol{k}})$
is related to the transition matrix $\mathbf{T}$ by \cite[Ch.~7.8]{ref_Sca}
\begin{equation}
\bar{\boldsymbol{S}}(\hat{\boldsymbol{r}},\hat{\boldsymbol{k}})
=
\sum_{\alpha,\beta}
\boldsymbol{\phi}_{\alpha}(\hat{\boldsymbol{r}})
T_{\alpha\beta}
\boldsymbol{\phi}_{\beta}^{*}(\hat{\boldsymbol{k}}).
\label{eq:appA_S_T_relation}
\end{equation}
Substituting \eqref{eq:appA_S_T_relation} into \eqref{eq:trcm_integral} gives
\begin{equation}
\int
\mathrm{d}\hat{\boldsymbol{k}}\,
\sum_{\alpha,\beta}
\boldsymbol{\phi}_{\alpha}(\hat{\boldsymbol{r}})
T_{\alpha\beta}
\boldsymbol{\phi}_{\beta}^{*}(\hat{\boldsymbol{k}})
\cdot
\boldsymbol{F}_{n}^{*}(-\hat{\boldsymbol{k}})
=
\sigma_n
\boldsymbol{F}_n(\hat{\boldsymbol{r}}).
\label{eq:appA_substitution}
\end{equation}
The left-hand side of \eqref{eq:appA_substitution} can be evaluated as
\begin{equation}
  \begin{split}
  &\sum_{\alpha,\beta}
  \boldsymbol{\phi}_{\alpha}(\hat{\boldsymbol{r}})
  T_{\alpha\beta}
  \int
  \mathrm{d}\hat{\boldsymbol{k}}\,
  \boldsymbol{\phi}_{\beta}^{*}(\hat{\boldsymbol{k}})
  \cdot
  \boldsymbol{F}_{n}^{*}(-\hat{\boldsymbol{k}})\\
  &\qquad\qquad =
  -\sqrt{Z_0}
  \sum_{\alpha,\beta}
  \boldsymbol{\phi}_{\alpha}(\hat{\boldsymbol{r}})
  T_{\alpha\beta}
  f_{n,\beta}^{*},
  \end{split}
\label{eq:appA_left_hand_side}
\end{equation}
where \eqref{eq:appA_conjugated_coefficient} has been used. The right-hand side of \eqref{eq:appA_substitution} is
\begin{equation}
\sigma_n \boldsymbol{F}_n(\hat{\boldsymbol{r}})
=
\sqrt{Z_0}
\sum_{\alpha}
\sigma_n f_{n,\alpha}
\boldsymbol{\phi}_{\alpha}(\hat{\boldsymbol{r}}).
\label{eq:appA_right_hand_side}
\end{equation}
Equating the expansion coefficients in the basis
$\{\boldsymbol{\phi}_{\alpha}\}$ yields
\begin{equation}
\mathbf{T}\mathbf{f}_{n}^{*}
=
-\sigma_n \mathbf{f}_n ,
\label{eq:appA_T_matrix_definition}
\end{equation}
which is the $\mathbf{T}$-matrix definition of time-reversal characteristic modes.

\section{Derivation of the Eigenvalue Bounds}
\label{App_B}

For a passive system, \eqref{eq:passivity_constraint} implies that, for every normalized modal vector $\mathbf{f}_n$,
\begin{equation}
  \label{eq:appB_passive_quadratic}
  \mathbf{f}_n^{\mathrm{T}} 
  \bigl(\mathbf{T}^{\mathrm{H}}\mathbf{T} + \Re\{\mathbf{T}\}\bigr)
  \mathbf{f}_n^{*} \le 0 .
\end{equation}
The first term in \eqref{eq:appB_passive_quadratic} can be evaluated using \eqref{eq:trcm_T_matrix} as
\begin{equation}
  \label{eq:appB_THT_term}
  \mathbf{f}_n^{\mathrm{T}}\mathbf{T}^{\mathrm{H}}\mathbf{T}\mathbf{f}_n^{*}
  =
  \bigl(\mathbf{T}\mathbf{f}_n^{*}\bigr)^{\mathrm{H}}
  \bigl(\mathbf{T}\mathbf{f}_n^{*}\bigr)
  =
  \sigma_n^{2}
\end{equation}
where the normalization $\mathbf{f}_n^{\mathrm{H}}\mathbf{f}_n=1$ has been used. Since $\Re\{\mathbf{T}\}$ is real symmetric, the second term becomes
\begin{equation}
  \label{eq:appB_ReT_term}
  \mathbf{f}_n^{\mathrm{T}}\Re\{\mathbf{T}\}\mathbf{f}_n^{*}
  =
  \Re\!\left\{
  \mathbf{f}_n^{\mathrm{T}}\mathbf{T}\mathbf{f}_n^{*}
  \right\}
  =
  -\sigma_n\,\Re\!\left\{\mathbf{f}_n^{\mathrm{T}}\mathbf{f}_n\right\}.
\end{equation}
Substituting \eqref{eq:appB_THT_term} and \eqref{eq:appB_ReT_term} into \eqref{eq:appB_passive_quadratic} gives
\begin{equation}
\sigma_n^2
-
\sigma_n
\Re\!\left\{\mathbf{f}_n^{\mathrm{T}}\mathbf{f}_n\right\}
\le 0 .
\label{eq:appB_sigma_ineq_first}
\end{equation}
Because $\sigma_n\ge 0$, this implies
\begin{equation}
  \label{eq:appB_sigma_bound}
  \sigma_n
  \le
  \Re\!\left\{\mathbf{f}_n^{\mathrm{T}}\mathbf{f}_n\right\}
  \le
  \left|\mathbf{f}_n^{\mathrm{T}}\mathbf{f}_n\right|
  \le
  \mathbf{f}_n^{\mathrm{H}}\mathbf{f}_n
  =
  1 .
\end{equation}
Therefore,
\begin{equation}
   \sigma_n \in [0,1].
\label{eq:appB_sigma_interval}
\end{equation}

Next, write
\begin{equation}
  \mathbf{f}_{n}^{\mathrm{T}}\mathbf{f}_n
  =
  b_n+\sigma_n-\mathrm{j}c_n,
  \qquad
  b_n\ge 0,\quad c_n\in \mathbb{R}.
\label{eq:appB_bn_cn_definition}
\end{equation}
Using the definition of $\tau_n$ in \eqref{eq:tau_definition}, we obtain
\begin{equation}
  \tau_n+\frac{1}{2}
  =
  \frac{1}{2}
  -
  \frac{\sigma_n}
  {\sigma_n+b_n+\mathrm{j}c_n}
  =
  \frac{b_n-\sigma_n+\mathrm{j}c_n}
  {2\left(\sigma_n+b_n+\mathrm{j}c_n\right)}.
\label{eq:appB_tau_shift}
\end{equation}
It follows that
\begin{equation}
  \left| \tau_n+\frac{1}{2} \right|
  =
  \frac{1}{2}
  \sqrt{
  \frac{
  c_n^{2}+\left(b_n-\sigma_n\right)^2
  }{
  c_n^{2}+\left(b_n+\sigma_n\right)^2
  }}
  \le
  \frac{1}{2}.
\label{eq:appB_tau_disk}
\end{equation}
Thus, $\tau_n$ is confined within the disk centered at $-1/2$ on the real axis with radius $1/2$.

\section{Solution of the Antilinear Eigenvalue Problem}
\label{App_C}

The antilinear eigenvalue problems encountered in this paper can be written in the following unified weighted form:
\begin{equation}
\mathbf{A}\mathbf{u}_n
=
\lambda_n \mathbf{B}\mathbf{u}_n^{*}
\qquad
\lambda_n \in \mathbb{R}
\label{eq:weighted_antilinear}
\end{equation}
where $\mathbf{A}$ is a complex symmetric matrix and $\mathbf{B}$ is a real weighting matrix. When $\mathbf{B}=\mathbf{1}$ (identity matrix), \eqref{eq:weighted_antilinear} reduces to the standard antilinear eigenvalue problem.

To convert \eqref{eq:weighted_antilinear} into a real-valued generalized eigenvalue problem, we decompose $\mathbf{A}$ and $\mathbf{u}_n$ into their real and imaginary parts:
\begin{equation}
\mathbf{A}
=
\mathbf{C}
+
\mathrm{j}\mathbf{D},
\qquad
\mathbf{u}_n
=
\mathbf{v}_n
+
\mathrm{j}\mathbf{w}_n .
\label{eq:real_imag_decomposition}
\end{equation}
Substituting \eqref{eq:real_imag_decomposition} into \eqref{eq:weighted_antilinear} gives
\begin{equation}
\left(\mathbf{C}+\mathrm{j}\mathbf{D}\right)
\left(\mathbf{v}_n+\mathrm{j}\mathbf{w}_n\right)
=
\lambda_n
\mathbf{B}
\left(\mathbf{v}_n-\mathrm{j}\mathbf{w}_n\right).
\label{eq:expanded_antilinear}
\end{equation}
Equating the real and imaginary parts yields
\begin{align}
\mathbf{C}\mathbf{v}_n
-
\mathbf{D}\mathbf{w}_n
&=
\lambda_n \mathbf{B}\mathbf{v}_n,
\label{eq:real_part_antilinear}
\\
\mathbf{D}\mathbf{v}_n
+
\mathbf{C}\mathbf{w}_n
&=
-\lambda_n \mathbf{B}\mathbf{w}_n .
\label{eq:imag_part_antilinear}
\end{align}
Introducing the block matrices
\begin{equation*}
\mathbf{L}
=
\begin{bmatrix}
\mathbf{C} & -\mathbf{D} \\
\mathbf{D} & \mathbf{C}
\end{bmatrix},
\qquad
\mathbf{M}
=
\begin{bmatrix}
\mathbf{B} & \mathbf{0} \\
\mathbf{0} & -\mathbf{B}
\end{bmatrix},
\qquad
\mathbf{x}_n
=
\begin{bmatrix}
\mathbf{v}_n \\
\mathbf{w}_n
\end{bmatrix}
\label{eq:real_generalized_matrices}
\end{equation*}
\eqref{eq:real_part_antilinear} and \eqref{eq:imag_part_antilinear} can be written compactly as
\begin{equation}
\mathbf{L}\mathbf{x}_n
=
\lambda_n \mathbf{M}\mathbf{x}_n .
\label{eq:real_generalized_eigenproblem}
\end{equation}
Thus, the original antilinear eigenvalue problem is transformed into a standard real-valued generalized eigenvalue problem, which can be solved using conventional eigensolvers.

It should be noted that \eqref{eq:real_generalized_eigenproblem} is an enlarged system whose dimension is twice that of the original problem. As a result, its eigenvalues always appear in positive--negative pairs. The two eigenvectors associated with each pair correspond to the same physical mode $\mathbf{u}_n$, differing only by a $90^{\circ}$ global phase shift. This pairing is a direct consequence of the real-valued reformulation and of doubling the system dimension from $N$ to $2N$. The choice between the positive and negative branches of $\lambda_n$ is therefore determined by the phase convention adopted in the modal definition. In this paper, when solving \eqref{eq:MoM_trcm}, the positive branch is selected because $\sigma_n$ is defined as a positive real number.

\end{appendices}

\end{document}